\documentclass[aps,prl,twocolumn,superscriptaddress]{revtex4}
\usepackage{graphicx}
\usepackage[colorlinks,citecolor=blue,urlcolor=blue, linkcolor=blue]{hyperref}
\usepackage{epstopdf}   
\usepackage{float}
\graphicspath{{figures//}}  
 
\begin{document}  
   
\title{High-energy radiation damage in zirconia: modeling results}       
\author{E. Zarkadoula} 
\affiliation{School of Physics and Astronomy, Queen Mary University of London, Mile End Road, London, E1 4NS, UK}
\affiliation{SEPnet, Queen Mary University of London, Mile End Road, London E1 4NS, UK}
\affiliation{Materials Science \& Technology Division, Oak Ridge National Laboratory, Oak Ridge, TN 37831, USA}
\author{R. Devanathan}
\affiliation{3 Nuclear Sciences Division, Pacific Northwest National Laboratory, Richland, WA 99352, USA}
\author{W. J. Weber}
\affiliation{Materials Science \& Technology Division, Oak Ridge National Laboratory, Oak Ridge, TN 37831, USA}
\affiliation{Department of Materials Science \& Engineering, University of Tennessee, Knoxville, TN 37996, USA}
\author{M. Seaton}
\affiliation{Scientific Computing Department, STFC Daresbury Laboratory, Keckwick Lane, Daresbury, Warrington, Cheshire, WA4 4AD, UK}
\author{I. T. Todorov}
\affiliation{Scientific Computing Department, STFC Daresbury Laboratory, Keckwick Lane, Daresbury, Warrington, Cheshire, WA4 4AD, UK}   
\author{K. Nordlund}
\affiliation{University of Helsinki, P.O. Box 43, FIN-00014 Helsinki, Finland} 
\author{M. T. Dove}
\affiliation{School of Physics and Astronomy, Queen Mary University of London, Mile End Road, London, E1 4NS, UK}
\author{K. Trachenko}
\affiliation{School of Physics and Astronomy, Queen Mary University of London, Mile End Road, London, E1 4NS, UK}
 \affiliation{SEPnet, Queen Mary University of London, Mile End Road, London E1 4NS, UK}

\begin{abstract}
Zirconia is viewed as a material of exceptional resistance to amorphization by radiation damage, and consequently proposed as a candidate to immobilize nuclear waste and serve as an inert nuclear fuel matrix. Here, we perform molecular dynamics simulations of radiation damage in zirconia in the range of 0.1-0.5 MeV energies with account of electronic energy losses. We find that the lack of amorphizability co-exists with a large number of point defects and their clusters. These, importantly, are largely isolated from each other and therefore represent a dilute damage that does not result in the loss of long-range structural coherence and amorphization. We document the nature of these defects in detail, including their sizes, distribution and morphology, and discuss practical implications of using zirconia in intense radiation environments.
\end{abstract}


\maketitle

Radiation effects have been finding increasing applications in various fields with notable examples of semiconductor and nuclear industries. In nuclear power applications, materials will be exposed to high dose irradiation coming from highly energetic products of fission and fusion. In these applications, the energy of emitted particles has a two-fold effect: on one the hand, this energy is converted into useful energy, by heating the material; on the other hand, this energy damages the material and degrades the properties important for the operation, including mechanical, thermal, transport and other properties. This is currently a central issue in the area of both fusion and fission energy generation \cite{stoneham, wasteforms3}. The latter in particular faces the problem of finding a material suitable for safe long-term encapsulation of nuclear waste \cite{wasteforms3,synroc2,gei1}. Candidate materials for safe encapsulation of nuclear waste, waste forms, need to be structurally stable and have low diffusion rates of radioactive ions to prevent polluting the environment. Waste forms often become amorphized by radiation damage from the nuclear waste, with most of amorphization produced by energetic heavy ions in collision cascades that consist of atoms displaced from their sites \cite{wasteforms3,synroc2,gei1,Ave98}. The diffusion can dramatically increase as a result of irradiation-induced amorphization \cite{gei1,gei2}. For this reason, the search for radiation-resistant waste forms has been on for several decades.

Zirconia, ZrO$_2$ (both in cubic and monoclinic form), stands out on the list of materials that are highly resistant to amorphization: both in-situ and ex-situ experiments such as X-ray diffraction, Rutherford backscaterring spectroscopy, transmission electron microscopy and so on indicated no loss of crystalline structure in bulk samples under bombardment with heavy MeV-energy ions and plutonium doping \cite{Sickafus,Costantini1,Costantini2,Sasajima,Weber1,Sickafus2,Sickafus3,Spino}. This was considered as evidence for the exceptional resistance to amorphization compared to other materials \cite{PRB_kostya_70,damage_evolution} and, combined with its ability to incorporate radioactive ions from nuclear waste including actinides \cite{burakov}, zirconia has been considered as a strong candidate material for inert fuel and nuclear waste matrices \cite{Sickafus,Costantini1,Costantini2,Sasajima,Weber1,Sickafus2,Sickafus3,Spino,fuels1,fuel2,wasteforms1,wasteforms2,wasteforms3,Lumpkin}.

An important question arises regarding what the high resistance to amorphization implies for the purposes of using zirconia as a waste form. The experimental probes above provide the information about the long-range order, and in many cases do not directly probe the nature of point defects and clusters smaller than several nanometers. Instead, these probes study the macroscopic consequences such as lattice swelling. This issue has been recently receiving increasing attention in the context of elucidating the {\it local}, as opposed to long-range, structure \cite{billinge}. In the case of zirconia, point defects and small clusters of point defects may not affect the Bragg peaks up to the smallest $k$-numbers, yet play an important role in defect-assisted diffusion processes involving the radioactive ions. Point defects have been seen in irradiated zirconia, although determination of their exact structure, abundance and distribution have been viewed as challenging \cite{Sickafus,Costantini1,Costantini2,Sasajima,Weber1,Sickafus2,Sickafus3,Spino}. In this paper, we address this question using molecular dynamics (MD) simulations, a method that provides access to detailed structural changes at the atomistic scale. We find that high-energy radiation damage creates unexpectedly large amount of damage. Importantly, this damage is contained in point defects and small clusters of point defects, and therefore does not constitute what is usually considered as amorphization in terms of the loss of long-range order, an insight that we additionally verify by direct calculation of the radial distribution function. In contrast to the present results, in some materials such as zirconium silicate (ZrSiO$_4$) \cite{JPCM_kostya} there is direct amorphization in the track and dense damage is produced. In our simulations we find that the defect atoms after the cascade relaxation represent dilute damage. We briefly discuss the implications of our findings for using zirconia as a waste form.

In this work, we perform MD simulations of radiation damage in cubic zirconia due to high energies in the 0.1--0.5 MeV range. To contain the damage due to these energies, we use system sizes with lengths of up to 130 nm and 150 millions of atoms. We use the DL$\_$POLY program, a general-purpose package designed for large-scale simulations \cite{dl1,dl2}. The combination of nearly-perfect scalability of the code with high--performance massive parallel computer facilities has recently set the stage for simulating systems of up to 1 billion atoms with realistic interatomic potentials \cite{eza}. We have used the interatomic potential that includes Buckingham pair interaction potentials and partial Coulomb charges \cite{pot-ram}. This potential stabilizes the cubic phase; we chose this potential to study defect production in a cubic ceramic phase. Experiments show \cite{balogh} that there is no phase transformation or grain restructuring during irradiation. Equilibrium U-O interaction was the same as Zr-O interaction. The system was equilibrated in the constant pressure ensemble at 300 K. U atom with 0.1 MeV energy was chosen as a primary recoil atom to correspond to the alpha-decay process. Higher velocities of U atoms of the range of 0.3--0.5 MeV correspond to heavy ion bombardment events. The initial position of the U atom is 30-50 \AA\ from the simulation box boundaries. The propagation of the recoil was followed in the constant energy and volume ensemble when there is no energy loss. When the electronic stopping mechanism in activated a friction term is included in the equation of motion. We are using variable timestep to account for faster atomic motion at the beginning of the cascade development and its gradual slowing down at later stages. To account for the highly non-equilibrium nature of radiation damage, the following modifications to the standard MD code were made. First, the MD box boundary layer of thickness of about 10 \AA\ was connected to a Langevin thermostat at 300 K to emulate the effect of energy dissipation into the sample. Second, we have accounted for the electronic energy losses, particularly important at high energies, by implementing the friction-type term in the equations of motion applied to particles above the certain cutoff energy $E_c$ (or velocity) \cite{duffy1,caro}. In metals, $E_c$ is often taken as approximately the double of the system's cohesion energy in order to differentiate ballistically moving atoms (with energy in excess of cohesion energy) from those oscillating. In insulators, it has been shown that the band gap governs the electronic energy losses during the radiation damage process \cite{emilio1,emilio2}, and we have accordingly set $E_c$ at twice the band gap in zirconia. The friction coefficient was calculation using SRIM tables \cite{srim}. 
Finally, the Buckingham potential was joined to a repulsive ZBL potential \cite{zbl} at short distances using a switching function \cite{zbl+potential_Kostya+Martin}. ZBL repulsion was also used for all pair interactions. The simulations were run on 3200--65000 parallel processors of UK's National Supercomputing Facilities, HECToR.

We have simulated recoils of 0.1, 0.3 and 0.5 MeV energies. Here, 0.1 MeV simulations are related to the recoil energy in Pu doping experiments where most of the structural damage comes from heavy recoils with approximately 0.1 MeV energy, whereas higher-energy events correspond to heavy ion bombardment experiments. We have simulated 0.1 MeV events with and without electronic energy loss whereas for higher energy events where a significant part of energy loss is due to electronic processes, the friction component was always on. For each simulation, we have simulated three randomly chosen directions of the recoil. The damage quantified below therefore refers to the average numbers and includes the standard deviation for each energy.

\begin{figure}
\begin{centering}
\includegraphics[width=\columnwidth]{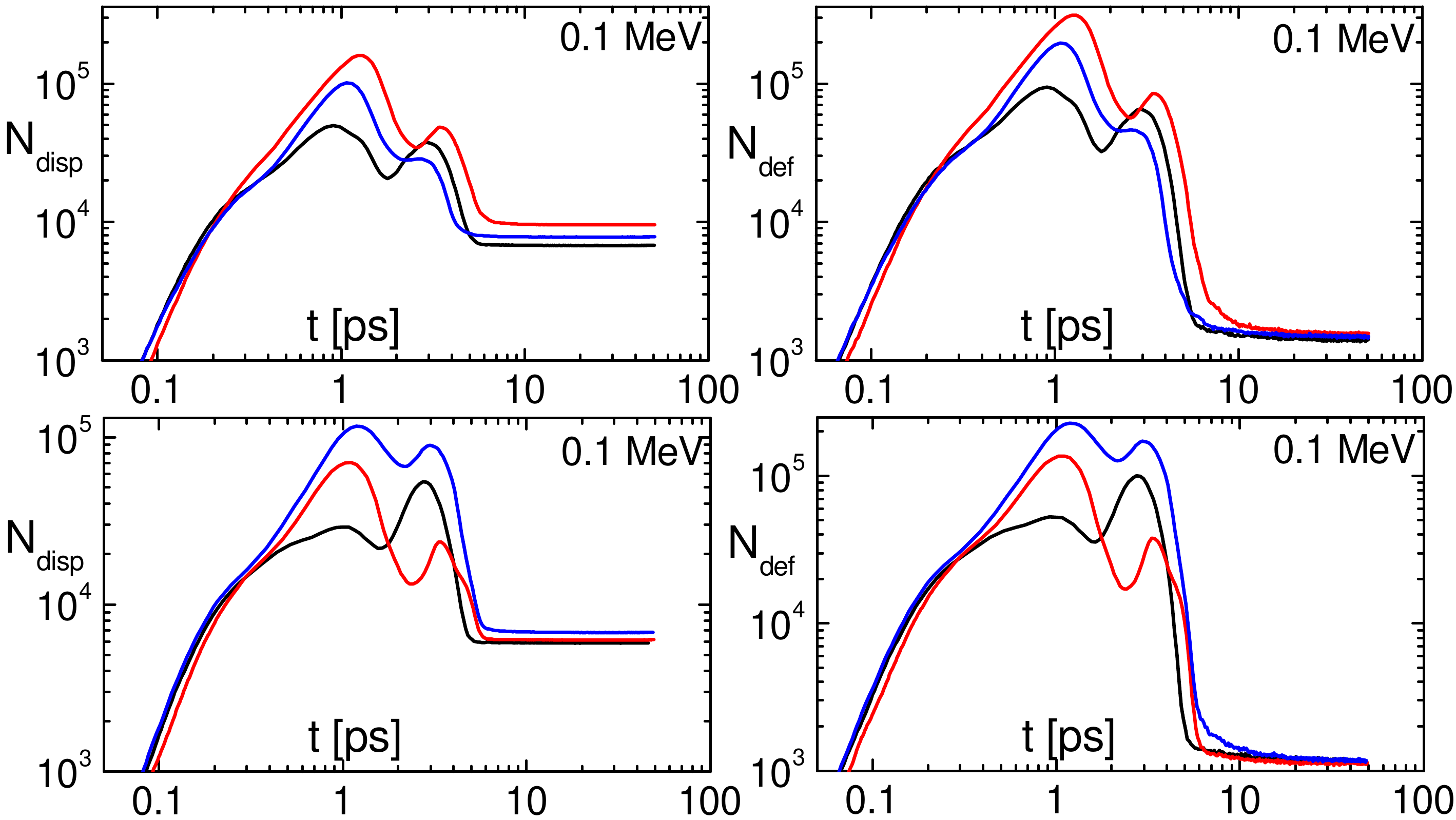}
\end{centering}
\caption{$N_{\mathrm{disp}}$ and $N_{\mathrm{def}}$ ($N_{\mathrm{def}}=N_{\mathrm{int}}+N_{\mathrm{vac}}$) from 0.1 MeV knock-on atoms without (top)and 0.1 MeV knock-on atoms with the effect of electronic energy loss switched on (bottom) for three directions of the recoil.
}
\label{fig1}
\end{figure}
 
\begin{figure}
\begin{centering} 
\includegraphics[width=\columnwidth]{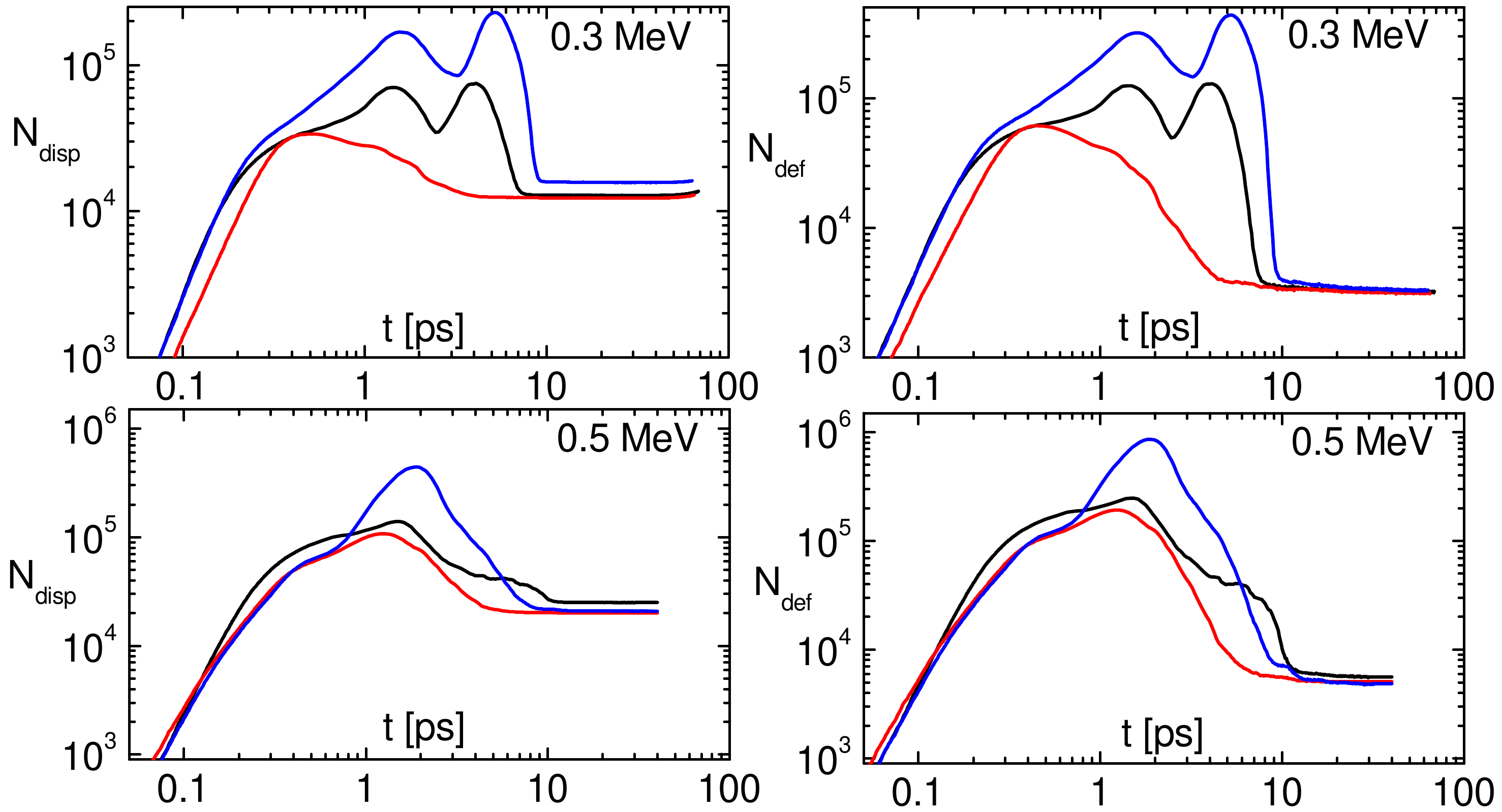}
\end{centering}
\caption{$N_{\mathrm{disp}}$ and $N_{\mathrm{def}}$ ($N_{\mathrm{def}}=N_{\mathrm{int}}+N_{\mathrm{vac}}$) from 0.3 MeV knock-on atoms (top) and 0.5 MeV knock-on atoms (bottom) for three directions of the recoil.
}
\label{fig2}
\end{figure}

\begin{table*}[thb]
\setlength{\tabcolsep}{0.2cm} 
\begin{tabular}{lcccc}
&  \multicolumn{2}{c}{\bf PEAK} & \multicolumn{2}{c}{\bf END} \\
   \hline\hline
 {PKA energy} & {\bf $N_{\mathrm{disp}}$}  &{\bf $N_{\mathrm{int}}$} (same for {\bf $N_{\mathrm{vac}}$}) & {\bf $N_{\mathrm{disp}}$}  &{\bf $N_{\mathrm{int}}$} (same for {\bf $N_{\mathrm{vac}}$})\\
  \hline
100 keV - no friction & 104,000 (32,000) & 96,000 (36,000) &  8,000 (800) & 750 (20) \\
100 keV - friction    & 72,000  (25,000) & 70,000 (26,000) &  6,000 (300) & 600 (10) \\
300 keV - friction    & 113,000 (60,000) & 105,000 (58,000) & 14,000 (1,000) & 1,600 (20) \\
500 keV - friction    & 230,000  (108,000) & 220,000 (107,000) & 22,000 (2,000) & 2,600 (100) \\ 
\hline\hline
\end{tabular}
\caption{$N_{\mathrm{disp}}$ and $N_{\mathrm{int}}$ ($N_{\mathrm{vac}}$), calculated using the sphere criterion, at the peak of the damage (1-2 ps) and at the end of the simulation. Standard error of the mean is shown in the brackets calculated over three events.}
\label{tab1} 
\end{table*}

We quantify damage production, evolution and recovery, and show the results in Figures \ref{fig1}-\ref{fig2}. We introduce two important quantities to describe the damage creation and recovery: $N_{\mathrm{disp}}$ and $N_\mathrm{def}$. $N_{\mathrm{disp}}$ accounts for the total displacements introduced in the system, i.e. is the number of atoms that have moved more than a cut--off distance ($d=0.75$ \AA) from their initial positions. To account for the atoms that recombine to crystalline positions, $N_\mathrm{def}$ is introduced. $N_\mathrm{def}$ reflects the recovery of structural damage as it corresponds to the sum of interstitials and vacancies. 
We use the sphere criterion for defect identification \cite{Nor97f}). If a particle, $p$, is located in the vicinity of a site, $s$, defined by a sphere with its center at this site and a radius, $d$, then the particle is a first-hand claimee of $s$, and the site is not vacant. Otherwise, the site is presumed vacant and the particle is presumed a general interstitial. If a site, $s$, is claimed and another particle, $p'$, is located within the sphere around it, then $p'$ becomes an interstitial associated with $s$. After all particles and all sites are considered, it is clear which sites are vacancies. Finally, for every claimed site, distances between the site and its first hand claimee and interstitials are compared and the particle with the shortest one becomes the real claimee. If a first-hand claimee of $s$ is not the real claimee it becomes an interstitial associated with $s$. At this stage it is clear which particles are interstitials. The sum of interstitials and vacancies gives the total number of defects in the simulated MD cell \cite{dlpoly_manual}. $d$ should generally be smaller than half of the closest interatomic separation, and is usually chosen not to account for typical thermal fluctuations of 0.2--0.3 \AA. With a certain choice of $d$, $N_{\mathrm{disp}}$ and $N_\mathrm{def}$ can be compared and agree with other methods of defect identification such as Wigner-Seitz analysis \cite{dynamic_annealing,Nor97f}.
 
Fig. \ref{fig1} shows $N_{\mathrm{disp}}$ and $N_\mathrm{def}$ for 100 keV cascades along three different knock--on directions, without (top) and with (bottom) the friction term. Fig. \ref{fig2} illustrates $N_{\mathrm{disp}}$ and $N_\mathrm{def}$ for 300 keV (top) and 500 keV  (bottom) cascades for the same directions of the U recoil atom. We observe large peaks of both $N_{\mathrm{disp}}$ and $N_\mathrm{def}$ for all simulated cascades, followed by the marked decrease and saturation after about 5--10 ps of simulation time. Peak and final, long-time, values of $N_{\mathrm{disp}}$ and number of interstitials (vacancies) $N_\mathrm{int}$ ($N_\mathrm{vac}$) are summarized in Table \ref{tab1}. We observe a substantial effect of the electronic friction on both
$N_{\mathrm{disp}}$ and $N_\mathrm{def}$, seen as a marked reduction of these numbers when the electronic friction is on. This effect originates from smaller energy available to produce both displaced atoms and surviving defects in the presence of electronic energy loss, and demonstrates the need to include electronic energy loss mechanisms in high-energy radiation damage simulations. We observe that the values of $N_{\mathrm{disp}}$ and $N_\mathrm{def}$ both at the peak and at the end of the cascade increase with increasing energy. Interestingly, we observed double peaks for cascades of 100 keV and 300 keV, which disappear for the higher energy cascades. This corresponds to the creation of subcascades for the lower energy cascades and more continuous damage morphology for the higher energy cascades \cite{eza}.

The physical origin of the large peaks is related to the deformation of the crystalline lattice around the collision cascade due to potential anharmonicity and associated expansion of the cascade structure, and was discussed in detail in our previous paper \cite{eza}. Here, we focus on the final values of $N_{\mathrm{disp}}$ and $N_\mathrm{def}$ at long times, and the latter in particular since it constitutes the remaining damage after the cascade relaxation. As far as long-term evolution is concerned, experiments of irradiated zirconia \cite{balogh} show that with time and temperature, defect density decreases due to annihilation at sinks.  Therefore, the primary damage state is the starting point to describe the evolution of the microstructure.

\begin{figure}
\begin{flushleft}
\includegraphics[width=\columnwidth]{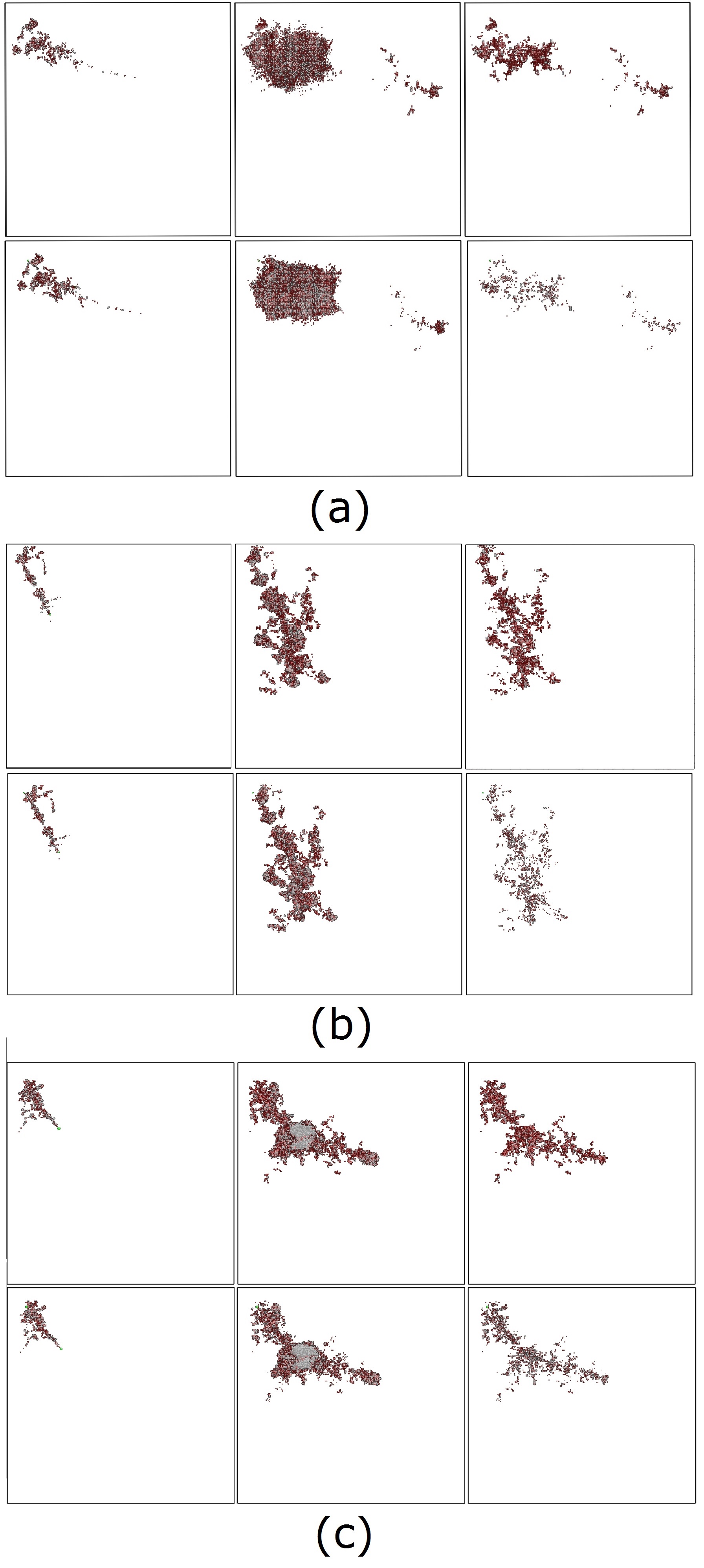}
\end{flushleft}
\caption{ Time frames of displaced and defect atoms for different recoil energy cascades with the effect of electronic energy loss switched on. The knock-on atom moves from the top left to the bottom right corner. Oxygen atoms are represented in red and zirconium atoms in gray. Top images represent the displaced atoms and the bottom images the defect atoms.(a) 0.1 ps, 1.1 ps and 50 ps in a 0.1 MeV collision cascade in a system with box length of 645 \AA, consisting of about 20 million atoms. Cascade size (maximal separation between any two defect atoms in the cascade) is about 600 \AA. (b) 0.1 ps, 0.45 ps and 64 ps for a collision cascade of 0.3 MeV collision cascade in a 70 million atoms system with box length of about 1000 \AA.  Cascade size is 800 \AA. (c) 0.1 ps, 1.2 ps and 17 ps for a 0.5 MeV collision cascade in a system of about 1300 \AA\ length, consisting of 150 million atoms. Cascade size is 1200 \AA.
}
\label{fig3}
\end{figure}

First, we observe large dynamic recovery of the induced damage, seen as the reduction of the final long-time $N_\mathrm{def}$ relative to $N_{\mathrm{disp}}$ in Figures \ref{fig1}-\ref{fig2}. The dynamic annealing is profound, and constitutes 80\%--99\% of damage recovery for different energies. This is consistent with earlier simulations of smaller 30 keV energy \cite{dynamic_annealing}, and is well illustrated in Fig. \ref{fig3} where we show both displaced and defect atoms for each simulated energy at various stages of damage propagation. In this figure, a typical cascade size, the maximal distance between defect atoms at the end of cascade propagation, is 600--1200 \AA. In 8 out of the 12 cascade simulations the U recoil is not identified as a "defect".

\begin{figure}
\begin{flushleft}
\includegraphics[width=\columnwidth]{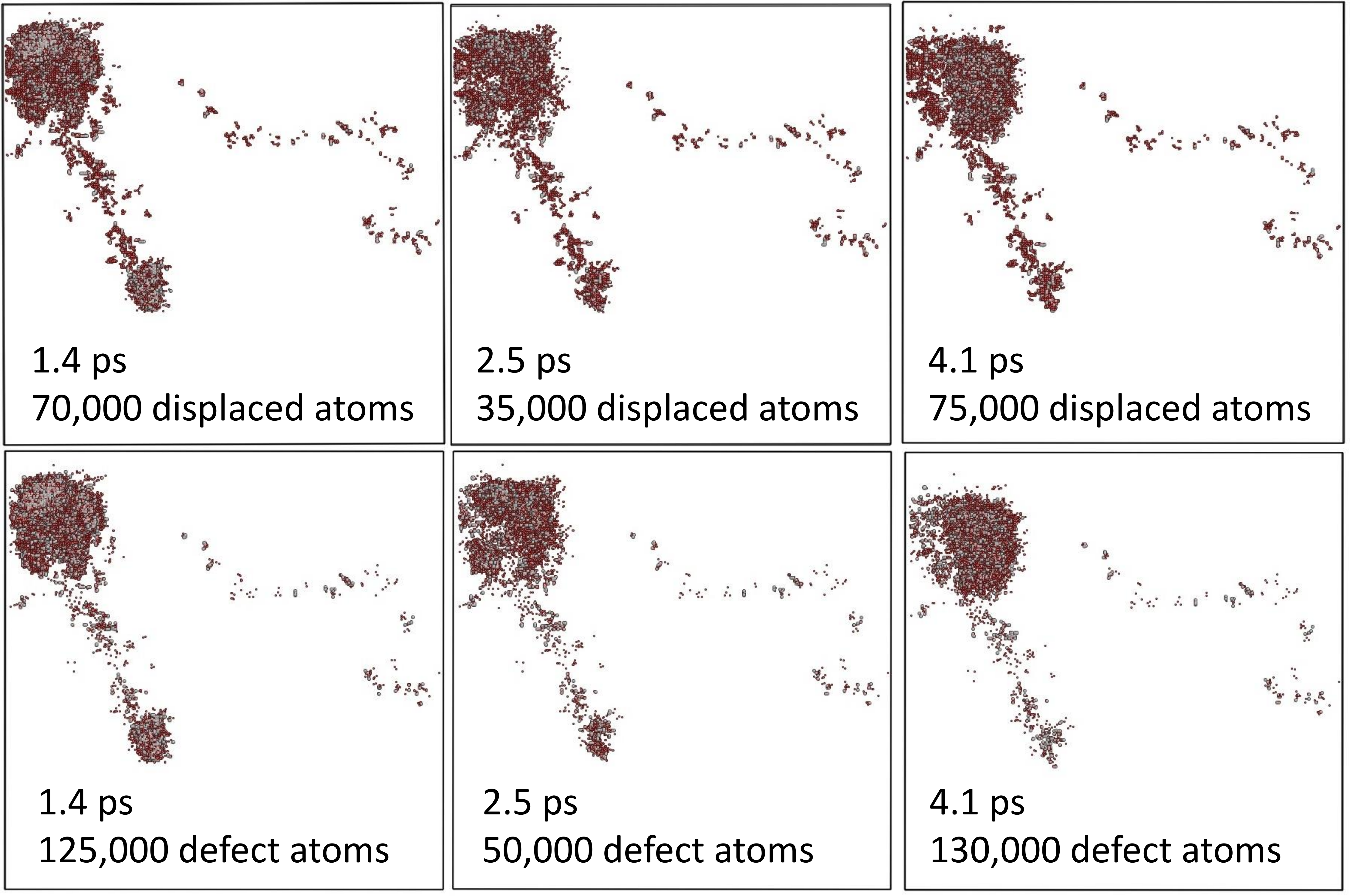}  
\end{flushleft}
\caption{Time frames of displaced and defect atoms for a 300 keV cascade where the effect of electronic energy loss is switched on. The knock-on atom moves from the top left to the bottom right corner. Oxygen atoms are represented in red and zirconium atoms in gray. The three frames shown are at 1.4 ps, 2.5 ps and 4.1 ps and correspond to the first peak, the following minimum and the second peak of the double-peak plot in the top of Fig.\ref{fig2}. The system consists of about 70 million atoms and has box length of about 1000 \AA.  Cascade size is about 800 \AA. The smaller pocket of damage reaches its maximal size and relaxes faster than the larger one on the top left corner of the simulation box. 
}  
\label{fig4}
\end{figure}

Fig.\ref{fig4} shows displaced and defect atoms in a 300 keV cascade at different time frames, which corresponds to the middle line (double-peak) shown in the top plots of Fig.\ref{fig2}. In contrast to the cascade shown in Fig.\ref{fig3}(b), which shows continuous shape of the damage and corresponds to the single peak line of Fig.\ref{fig2}, we see a resolvable smaller subcascade at the bottom. These two pockets expand and relax with time difference of about 2.5 ps. The first frame is at 1.4 ps where the bottom subcascade reaches its maximal size, corresponding to the first peak shown in Fig.\ref{fig2}. In the second frame (2.5 ps) the smaller pocket reduces in size, followed by the larger pocket (left top corner) reaching its maximal size at 4.1 ps in the third frame. This is confirmed by counting the atoms in both subcascades at different times. The dilute damage on the right of the cascade in Fig. 4 is due to channelling in this event.

The key to reconciling the exceptional radiation tolerance of ZrO$_2$ on experimental basis and the large number of defect atoms comes from realization that the damage at the end of the simulation time is very dilute, and mostly consists of isolated point defects and small disjoint clusters. In this case, X-ray probes, TEM and other non-local experimental probes do not detect amorphization as the loss of the long--range order, and hence consider zirconia as highly resistant despite the presence of the large number of local defects.

\begin{figure}
\begin{flushleft}
\includegraphics[width=\columnwidth]{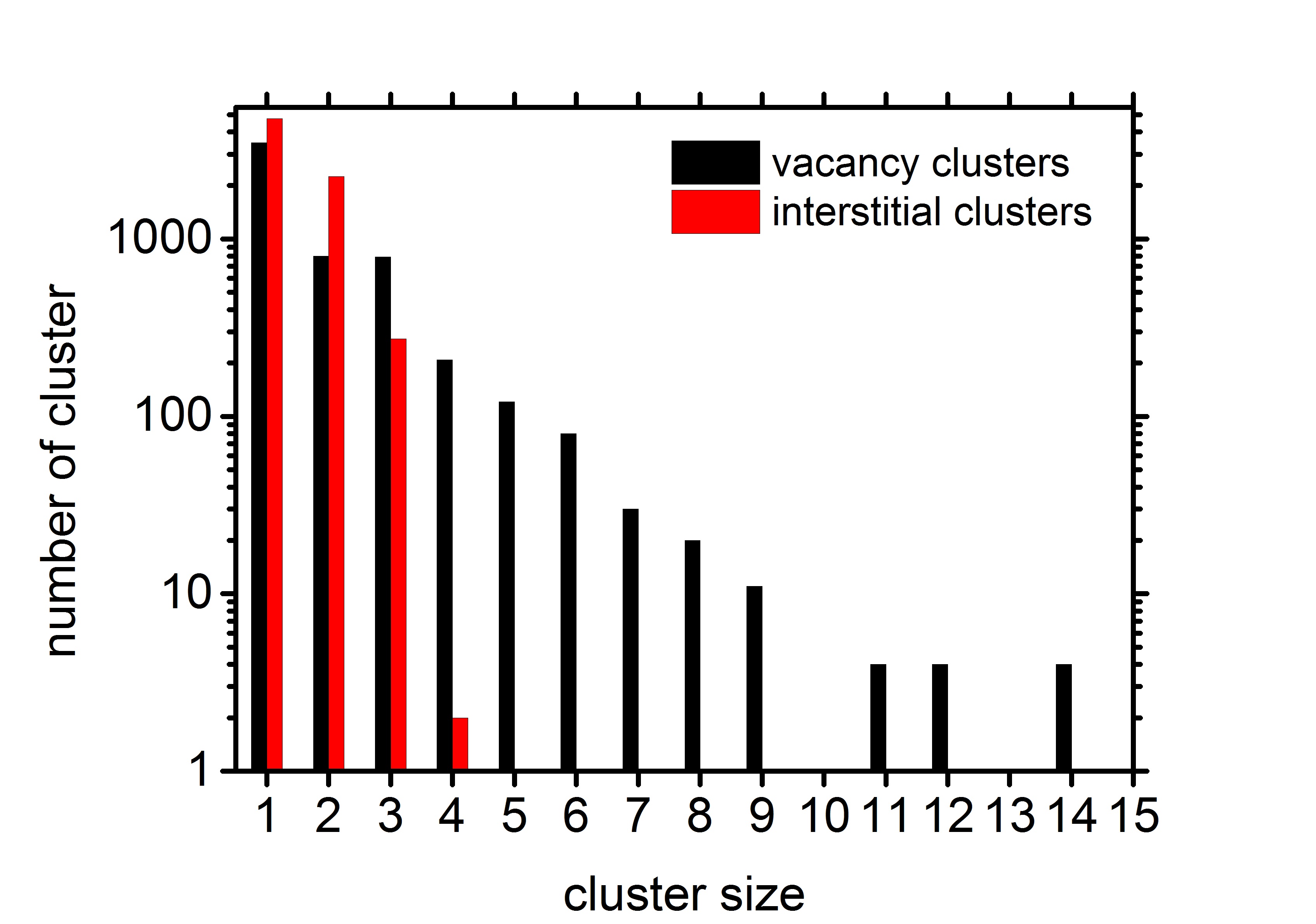}
\end{flushleft}
\caption{ Distribution of cluster sizes for all cascades performed.}
\label{fig5}
\end{figure}

First, we support our proposal by the detailed analysis of defect atoms and the cascade morphology. In Table \ref{tab2}, we summarize how $N_\mathrm{def}$ partition in clusters of different sizes. As discussed later, we find that across all cascades simulated, the majority of the defects are isolated, most of which are O vacancies. In Fig. \ref{fig5}, we show the distribution of cluster sizes, and similarly find that most of the damage resides in isolated point defects and small clusters. Notably, we find that vacancy clusters to be appreciably larger than interstitial clusters (see Table \ref{tab2} and Fig. \ref{fig5}), the point to which we return below.

\begin{table*}[t] 
\setlength{\tabcolsep}{0.2cm}
\begin{tabular}{p{4cm} l p{2.0cm} p{2.5cm} p{2.5cm} p{2.5cm} p{2.5cm} p{2.5cm}}
{\bf Property}&	      {\bf Cascade}&  {\bf 100keV fr}&	{\bf 100keV no fr}&	{\bf 300keV}&	{\bf 500keV}& \\ [2ex]
Surviving Zr vac&	1&  103&	141&	252&	390&\\
(same for Zr int)&	    2&	105&	117&	257&	465&\\
\hspace{2cm}        &  	3&	103&    123&	269&	391&\\ [2ex]
	
Surviving O vac&	1&	242&	338&	742&	1105&\\
(same for O int)&	    2&	266&	310&	741&	1219&\\
\hspace{2cm}    &       3&	257&	357&	759&	1035&\\ [2ex]
					
Surviving defects&	     1&	690&	958&	1988&	2990&\\
(vac and int of Zr and O)&	 2&	742&	854&	1996&	3368&\\
\hspace{2cm}        &    3&	720&	960&	2056&	2852&\\ [2ex]
					
Longest distance& 	1&	585.88&	453.90&	830.57&	 1177.85&\\
between defects (\AA)&	    2&	380.08&	440.39&	1055.44& 1262.62&\\
\hspace{2cm}       &     3&	365.01&	559.40&	1019.81& 1240.10&\\ [2ex]

Zr int-vac distance (\AA)	&	1	&	12.68	&	12.66	&	12.23	&	12.02	&\\
\hspace{2cm}	&	2	&	12.07	&	12.35	&	11.69	&	12.37	&\\
\hspace{2cm}	&	3	&	12.84	&	11.93	&	12.29	&	12.14	&\\ [2ex]
											
O int-vac distance (\AA)	&	1	&	12.25	&	11.31	&	10.30	&	10.41	&\\
\hspace{2cm}	&	2	&	11.16	&	10.82	&	10.47	&	10.86	&\\
\hspace{2cm}	&	3	&	11.33	&	10.91	&	10.39	&	10.83	&\\ [2ex]

Fraction of O vac	&	1	&	0.70	&	0.71	&	0.75	&	0.74	&\\
\hspace{2cm}	&	2	&	0.72	&	0.73	&	0.74	&	0.72	&\\
\hspace{2cm}	&	3	&	0.71	&	0.74	&	0.74	&	0.73	&\\ [2ex]
											
Fraction of O int	&	1	&	0.70	&	0.71	&	0.75	&	0.74	&\\
\hspace{2cm}	&	2	&	0.72	&	0.73	&	0.74	&	0.72	&\\
\hspace{2cm}	&	3	&	0.72	&	0.74	&	0.74	&	0.73	&\\ [2ex]
											
Fraction isolated vac	&	1	&	0.26	&	0.26	&	0.38	&	0.36	&\\
\hspace{2cm}	&	2	&	0.33	&	0.30	&	0.38	&	0.32	&\\
\hspace{2cm}	&	3	&	0.32	&	0.36	&	0.38	&	0.36	&\\ [2ex]
											
Fraction isolated int	&	1	&	0.42	&	0.42	&	0.50	&	0.48	&\\
\hspace{2cm}	&	2	&	0.46	&	0.44	&	0.49	&	0.47	&\\
\hspace{2cm}	&	3	&	0.45	&	0.50	&	0.48	&	0.47	&\\ [2ex]
											
Frac O in isolated vac	&	1	&	0.91	&	0.94	&	0.96	&	0.95	&\\
\hspace{2cm}	&	2	&	0.91	&	0.92	&	0.94	&	0.94	&\\
\hspace{2cm}	&	3	&	0.93	&	0.94	&	0.93	&	0.92	&\\ [2ex]
											
Frac O in isolated int	&	1	&	0.96	&	0.95	&	0.96	&	0.97	&\\
\hspace{2cm}	&	2	&	0.94	&	0.98	&	0.97	&	0.95	&\\
\hspace{2cm}	&	3	&	0.96	&	0.96	&	0.97	&	0.95	&\\ [2ex]
											
Size of largest vac cluster	&	1	&	11	&	13	&	12	&	14	&\\
\hspace{2cm}            	&	2	&	9	&	9	&	12	&	17	&\\
\hspace{2cm}	            &	3	&	14	&	21	&	11	&	14	&\\ [2ex]
											
Size of largest int cluster	&	1	&	5	&	3	&	3	&	3	&\\
\hspace{2cm}	            &	2	&	3	&	4	&	3	&	4	&\\
\hspace{2cm}            	&	3	&	3	&	3	&	4	&	3	&\\ [2ex]

\hline
\end{tabular}

\caption{Defect analysis for different recoil energy cascades, for three different directions of the recoil. fr stands for friction, vac for vacancies, int for interstitials }
\label{tab2}

\end{table*}

\begin{figure}
\begin{flushleft}
\includegraphics[width=\columnwidth]{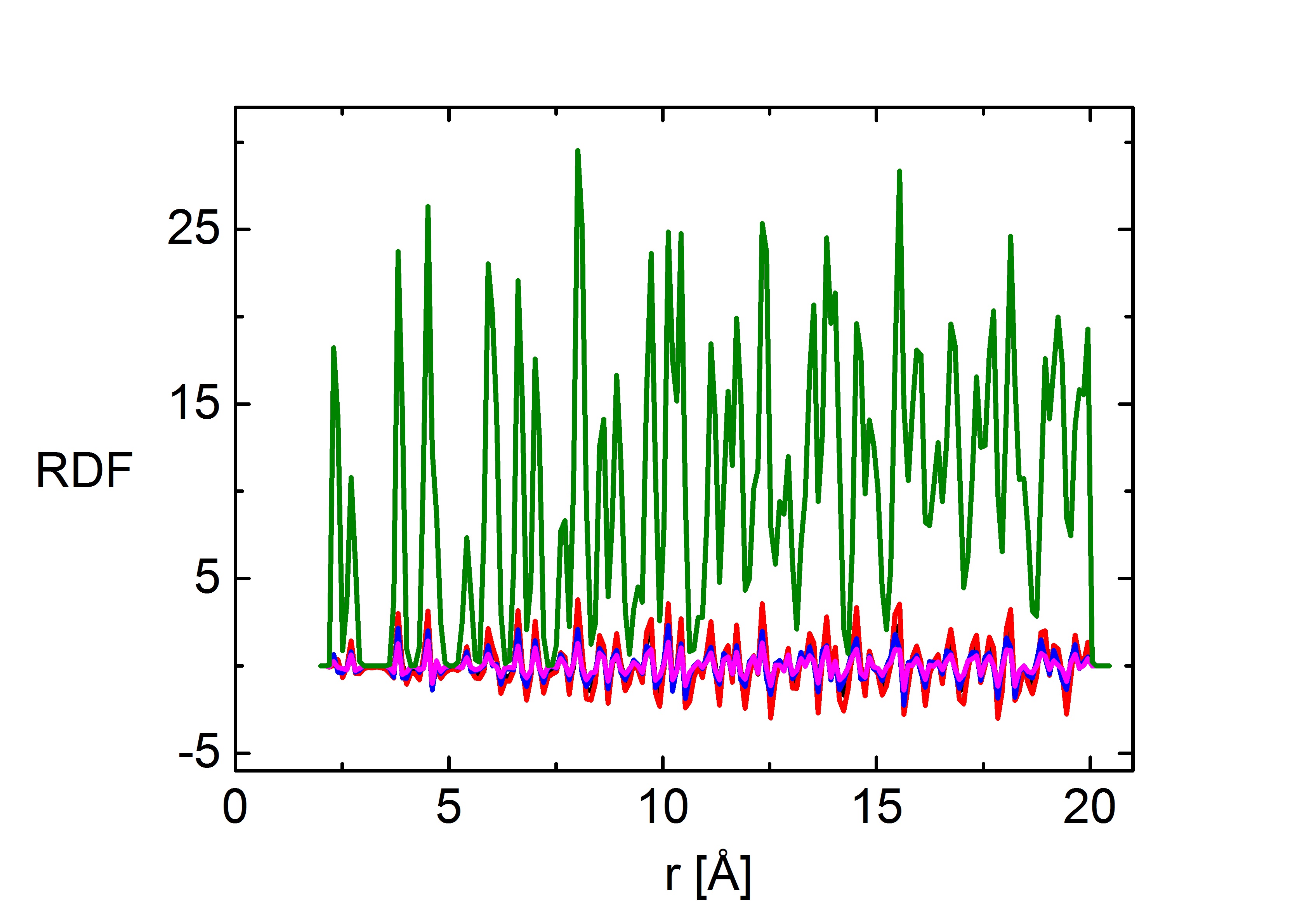}
\end{flushleft}
\caption{$t(r)$ calculated for the crystalline system and the differences between $t(r)$ for the crystalline structure and $t(r)$ for the damaged structures for four collision cascades.}
\label{fig6}  
\end{figure}

Second, we calculate the radial distribution function (RDF) over atoms in four simulated collision cascades. We define the radius of gyration of the collision cascade as $R_\mathrm{G}=\sum\Delta{\bf r}/N$, where ${\bf\Delta r}$ is the distance between interstitials and the centre of gyration. We calculate RDF as a histogram of separations between all atoms located within the sphere of radius $R_\mathrm{G}$ centered at ${\bf r}_\mathrm{C}=\sum {\bf r_i}/N$, where ${\bf r_i}$ are positions of identified interstitial defect atoms $i$ and $N$ is their number. In Fig. \ref{fig6} we show $t(r)=g(r)\cdot r$, where $g(r)$ is  normalized to the value 1 for large distances, calculated for the crystalline structure and the difference of $t(r)$ between the crystalline structure and four collision cascades. Fig. \ref{fig6} shows the near-identity of RDFs between the damaged and crystalline structure including, importantly, the presence of peaks beyond the short- and medium-range order. This is in contrast to the disappearance of peaks beyond the medium-range order in systems such as SiO$_2$, TiO$_2$, ZrSiO$_4$ and so on where in-cascade amorphization is observed \cite{kos73,farnan1,farnan2,JPCM_kostya}.

We now give more details about the nature of radiation damage in ZrO$_2$ presented in Table \ref{tab2}. We perform the analysis excluding vacancy-interstitial (V-I) pairs of the same species (both cation or both O) that lie within 3 \AA\ of each other under the assumption that they will quickly annihilate with each other. We define clusters based on defects being within 3 \AA\ of a defect, i.e for vacancy - vacancy (V-V) distance and interstitial - interstitial (I-I) distance less 3 than \AA. The number of surviving defects increases with increasing energy with an almost linear dependence. Our numbers of interstitials as a function of energy for the friction case give an exponent of about 0.9.  If $n$ is the number of interstitials and $E$ is the PKA energy, then according to our data $(n_1/n_2) = (E_1/E_2)^{0.9}$. The defect production in previous work in metals and ceramics in simulations of PKA energy up to 150 keV  \cite{martin2011,st-rev}, where the electronic effects have not been taken into account, shows linear dependence on the PKA energy.

The longest separation between defects in the cascade increases with increasing energy and is also larger if no friction is applied, reflecting that including electronic stopping  in high energy cascades results in smaller cascade size.  This is also demonstrated by the radius of gyration at the end of the cascades, with values of 53, 91, 141, 210 \AA\ averaged over three events for 100 keV cascades with electronic stopping, 100 keV cascades without the friction term applied, 300 keV and 500 keV cascades respectively. There is no obvious trend in the average distance between vacancies and nearest interstitials on the same sublattice with increasing energy or inclusion of friction.  However, the Zr V-I separation is slightly larger than the O V-I separation, with the standard deviation of this distance being larger for the O sublattice.  Therefore O defects can be found with large V-I separations. The fraction of vacancies and interstitials that are associated with the anion sublattice is 0.71 to 0.74 regardless of energy or friction. 30-38\% of vacancies are isolated and this percent does not vary significantly with energy or friction. 44-49\% of interstitials are isolated and again there is not much variation with energy or friction. Interstitials are more likely to be isolated than vacancies. Roughly one half of interstitials and on third of vacancies are isolated.  

Consistent with the above, interstitial cluster sizes are smaller than vacancy cluster sizes.  The largest interstitial cluster, consisting of both Zr and O, size is consistently between 3 and 4 regardless of energy or friction. Vacancy clusters are larger and the size of the largest cluster increases with energy and also if friction is ignored. The largest vacancy cluster is between 11 and 21 vacancies.  These vacancy clusters are either 50\% O (Zr-O clusters) or fairly close to the stoichiometric ratio (O-Zr-O vacancy clusters). Fraction of oxygen among isolated vacancies is 0.92 to 0.94 and among isolated interstitials is 0.95 to 0.97.  This seems independent of the energy and electronic stopping effect. Isolated defects are predominantly (almost entirely) on the oxygen sublattice.

We have brought insight into the primary damage state, by going beyond previous work in taking the electronic effects into account through the friction term. This information can serve as input for mesoscale models that incorporate grain boundaries and sinks and propagate the system much farther in time than possible with MD simulations.  The results of such modeling can be directly compared to experiments and will be the subject of future study. Even though zirconia does not amorphize in a sense of losing long-range order as most of the systems do \cite{JPCM_kostya}, a large number of point defects and their clusters found here may play an important role in long-term evolution of the damage \cite{uber} and increased diffusion of radioactive species in particular. Especially relevant in this context is the larger size of the vacancy clusters (9--21, see Table \ref{tab2} and Fig. \ref{fig5}) as compared to the size of interstitial clusters (3--5, see Table \ref{tab2} and Fig. \ref{fig5}). Consistent with experimental results \cite{Sickafus2,Sasajima,Weber1,Spino}, these can provide fast diffusion pathways for encapsulated radioactive species, but also play a role in the nucleation and growth of bubbles affecting the overall performance of the waste form. Experiments also show that defect clusters can also serve as sinks for point defects \cite{balogh}. Similar effects can be relevant in materials such as the widely used nuclear fuels UO$_2$ that are similar to zirconia in terms of structure and bonding. 

In summary, we have found that a large number of point defects and their clusters co-exist with long-range structural coherence in irradiated zirconia. These defect structures are largely disjoint from each other and therefore represent a dilute damage that does not result in the loss of long-range structural coherence and amorphization. At the same time, long-time evolution of these defects may have important implications for using zirconia in intense radiation environments. 

This work made use of the facilities of HECToR, via the Materials Chemistry Consortium, funded by EPSRC (EP/F067496). RD and WJW were supported by the U.S. Department of Energy, Office of Basic Energy Sciences, Division of Materials Sciences and Engineering.

\end{document}